\documentclass{article}
\usepackage{ijcai19}

\usepackage{times}
\usepackage{soul}
\usepackage{url}
\usepackage[hidelinks]{hyperref}
\usepackage[utf8]{inputenc}
\usepackage[small]{caption}
\usepackage{graphicx}
\usepackage{amsmath}
\usepackage{booktabs}
\usepackage{algorithm}
\usepackage{algorithmic}
\usepackage{subfigure} % for two graphs in one line
\usepackage{amsfonts} % blackboard math symbols
\usepackage{amsmath}
\usepackage{caption}
\usepackage{bm} % use \bm to bold the text in $*$, such as $a = b + c$
\usepackage{color}

\title{Learning Multi-agent Communication under Limited-bandwidth Restriction \\ for Internet Packet Routing}

\author{
    anonymous authors % Sarit Kraus
    \affiliations
    anonymous affiliations
    \emails
    anonymous emails
}

\author{
Hangyu Mao $^1$ \and
Zhibo Gong $^2$ \and
Zhengchao Zhang $^1$ \and
Zhen Xiao $^1$ \And
Yan Ni $^{1,3}$
\affiliations
$^1$ Peking University \\
$^2$ Huawei Technologies Co., Ltd. \\
$^3$ Microsoft
\emails
\{hy.mao, zhengchaozhang, xiaozhen, yanni\}@pku.edu.cn,
gongzhibo@huawei.com % leckie.dove@gmail.com
}
\begin{document}

\maketitle

\begin{abstract}
  Communication is an important factor for the big multi-agent world to stay organized and productive. Recently, the AI community has applied the Deep Reinforcement Learning (DRL) to learn the communication strategy and the control policy for multiple agents. However, when implementing the communication for real-world multi-agent applications, there is a more practical \emph{\textbf{limited-bandwidth}} restriction, which has been largely ignored by the existing DRL-based methods. Specifically, agents trained by most previous methods keep sending messages incessantly in every control cycle; due to emitting too many messages, these methods are unsuitable to be applied to the real-world systems that have a limited bandwidth to transmit the messages. To handle this problem, we propose a gating mechanism to adaptively prune unprofitable messages. Results show that the gating mechanism can prune more than 80\% messages with little damage to the performance. Moreover, our method outperforms several state-of-the-art DRL-based and rule-based methods by a large margin in both the real-world packet routing tasks and four benchmark tasks.
\end{abstract}

\section{Introduction}
Communication is an essential human intelligence, while Reinforcement Learning (RL) and especially the Deep Neural Network (DNN) based Reinforcement Learning (DRL) are emergent techniques that can achieve artificial intelligence. Recently, inspired by the communication among humans, DRL-based methods have been successfully applied to learn the communication among multiple intelligent agents. % Specifically, in cooperative multi-agent systems, communication makes sure that multiple agents can cooperate towards shared goals. Particularly, in partially observable environments, the agents can only observe different parts of the environment, communication allows the agents to share their observations to form a better representation of the environment state.

However, it remains an open question to apply these methods to the real-world multi-agent systems, because real-world applications usually impose many constraints on communication, such as the bandwidth limitation for
transmitting messages, the random delay of messages, and even the protection of private messages. Only by resolving these constraints can we develop practical communication strategy. % In order to develop practical communication strategy, it is necessary to consider these constraints.

In this paper, we focus on resolving the \textbf{\emph{limited-bandwidth restriction}} in multi-agent communication. We then try to apply our method to the real-world packet routing systems, because packet routing is not only a representative task with the above property, but also one of the most essential and critical tasks on the Internet. % by adopting DRL technique

Formally, we define the limited-bandwidth restriction as follows: the bandwidth (or more generally, the resource) for transmitting the communication messages is limited, therefore the agents should generate as few messages as possible \emph{on the premise of maintaining the performance}.

We are interested in the limited-bandwidth restriction due to two reasons. On the one hand, it is ubiquitous in real systems. For example, in the packet routing systems, the link has a limited transmission capacity; in the Internet of Things, the sensor has a limited battery capacity. Once we figure out a principled method to address this problem, many fields can benefit from this work. On the other hand, this problem has been largely ignored by the existing DRL-based methods. There is a great need to devote attention to this problem. %  On the one hand,

We take two steps to address this problem. Firstly, we aggregate the merits of the existing methods to form a basic model named \emph{Actor-Critic with Message Learning} (ACML). However, the proposed ACML is still not practical because it does not change the communication pattern of the existing methods. That is to say, ACML keeps sending messages incessantly in \emph{\textbf{every}} control cycle, regardless whether the message is beneficial enough for the whole agent team. Secondly, we extend ACML with a \textbf{G}ating mechanism to design a more flexible and practical \textbf{G}ACML model. The gating mechanism is trained based on a novel auxiliary task, which tries to open the gate to encourage communication when the message is beneficial enough to the whole agent team, and close the gate to discourage communication otherwise. As a result, after the gating mechanism is trained well, it can prune unprofitable messages adaptively to control the message quantity around a desired threshold. Consequently, GACML is applicable to real-world systems with limited-bandwidth restriction.

We evaluate our method in the real-world packet routing and benchmark tasks. It outperforms several state-of-the-art DRL-based and rule-based methods by a large margin. Furthermore, the proposed gating mechanism can prune more than 80\% messages with little damage to the performance.

\section{Background}
\textbf{DEC-POMDP.} We consider a partially observable cooperative multi-agent setting that can be formulated as DEC-POMDP \cite{bernstein2002complexity}. It is formally defined as a tuple $\langle N,S,\vec{A},T,R,\vec{O},Z,\gamma \rangle$, where $N$ is the number of agents; $S$ is the set of state $s$; $\vec{A}=[A_1, ..., A_N]$ represents the set of \emph{joint action} $\vec{a}$, and $A_i$ is the set of \emph{local action} $a_i$ that agent $i$ can take; $T(s'|s,\vec{a}): S \times \vec{A} \times S \rightarrow [0,1]$ represents the state transition function; $R: S \times \vec{A} \times S \rightarrow \mathbb{R}$ is the reward function shared by all agents; $\vec{O} = [O_1, ..., O_N]$ is the set of \emph{joint observation} $\vec{o}$ controlled by the observation function $Z: S \times \vec{A} \rightarrow \vec{O}$; $\gamma \in [0,1]$ is the \emph{discount factor}.

In a given state $s$, agent $i$ can only observe an observation $o_i \in s$, and each agent takes an action $a_i$ based on its own observation $o_i$, resulting in a new state $s'$ and a shared reward $r$. The agents try to learn a policy $\pi_{i}(a_i|o_i): O_i \times A_i \rightarrow [0,1]$ that can maximize $\mathbb{E}[G]$ where $G$ is the \emph{discount return} defined as $G = \sum_{t=0}^{H} \gamma^{t} r^{t}$, and $H$ is the time horizon. In practice, we map the observation history rather than the current observation to an action (namely, $o_i$ represents the observation history of agent $i$ in the rest of the paper).

\textbf{Reinforcement Learning (RL).} RL \cite{sutton1998introduction} is generally used to solve special DEC-POMDP problems where $N=1$. In practice, we usually define the Q-value function as $Q^{\pi}(s,a) = \mathbb{E}_{\pi}[G|S=s,A=a]$, then the optimal policy can be derived by $\pi^{*} = \arg\max_{\pi} Q^{\pi}(s,a)$.

Deterministic Policy Gradient (DPG) \cite{silver2014deterministic} is a special actor-critic algorithm where the actor adopts a deterministic policy $\mu_{\theta}: S \rightarrow A$ and the action space $A$ is continuous. Deep DPG (DDPG) \cite{lillicrap2015continuous} applies DNN $\mu_{\theta}(s)$ and $Q(s,a;w)$ to approximate the actor and the critic, respectively. DDPG is an off-policy method. It adopts \emph{target network} and \emph{experience replay} to stabilize training and to improve data efficiency. Specifically, the critic's parameters $w$ and the actor's parameters $\theta$ are updated based on: % the following equations:
\begin{eqnarray}
    \delta =& \hspace{-0.5em} r + \gamma Q(s',a';w^{-})|_{a'=\mu_{\theta^{-}}(s')} - Q(s,a;w) \label{equ:DPG1} \\
    L(w) =& \hspace{-11.5em} \mathbb{E}_{(s,a,r,s') \sim D}[\delta^{2}] \label{equ:DPG2} \\
    \nabla_{\theta}J(\theta) =& \hspace{-1.9em} \mathbb{E}_{s \sim D}[\nabla_{\theta}\mu_\theta(s) * \nabla_{a}Q(s,a;w)|_{a=\mu_{\theta}(s)}] \label{equ:DPG3}
\end{eqnarray}
where $D$ is the replay buffer containing recent experience tuples $(s,a,r,s')$; $Q(s,a;w^{-})$ and $\mu_{\theta^{-}}(s)$ are the target networks whose parameters $w^{-}$ and $\theta^{-}$ are periodically updated by copying $w$ and $\theta$. A merit of actor-critic algorithms is that the critic is only used during training, while only the actor is needed during execution. Due to this merit, we only need to prune messages among the actors in our GACML. %, because critic will be naturally removed during execution.

\section{Related Work} \label{sec:RelatedWork}
\textbf{Traditional Communication Model.} The communication models have been widely studied in the RL community, e.g., MTDP-COM \cite{pynadath2002communicative} and DEC-POMDP-COM \cite{goldman2004decentralized}. However, traditional methods usually either predefine the communication message \cite{wu2011online} or optimize the communication message for a predefined control policy \cite{roth2005reasoning}, which are inapplicable to the real-world multi-agent systems. Previous studies also try to address the limited-bandwidth restriction by pruning the messages \cite{roth2005reasoning,becker2009analyzing,wu2011online}. A general method is the Value of Communication (VoC) \cite{becker2009analyzing}. VoC measures the difference between the expected values of communicating and remaining silent. Communication is a better choice when VoC is larger than zero. Nevertheless, calculating VoC requires the knowledge of the environment, which is not easy to get since multi-agent systems are usually too complex. In contrast, we focus on model-free learning.

\textbf{Deep Communication Model.} Recently, combining DNN with RL, DRL-based communication models have been explored in model-free setting, such as CommNet \cite{sukhbaatar2016learning}, DIAL \cite{foerster2016learning}, BiCNet \cite{peng2017multiagent}, AMP \cite{peng2018learning} and \cite{mao2017accnet,mordatch2018emergence,mao2018modelling}. They adopt a DNN with hard-coded structure to represent the policy, and the policy takes as input the messages from \emph{all} agents to generate \emph{a single} control action. Thus, the agents have to keep sending messages incessantly in every control cycle, without alternatives to reduce the message quantity. Due to emitting too many messages, they are inflexible to be applied to multi-agent systems with the limited-bandwidth restriction. %  and many other methods \cite{mao2017accnet,kong2017revisiting}   the real-world

Methods for addressing the limited-bandwidth restriction can be roughly divided into two types. The simple type does not generate messages at all. For example,  MADDPG \cite{lowe2017multi} and COMA \cite{foerster2017counterfactual} adopt an independent DNN to generate the policy for each individual agent. Because the policy is independent of other agents' information, these methods can generate control action without any communication. Therefore, they are applicable in tasks with strict limit-bandwidth restriction. However, as the policies do not exchange messages, these methods suffer from the partially observable problem. Obviously, it is impossible to figure out the best control policies based on partial information. % , and other methods \cite{gupta2017cooperative,rashid2018qmix} The simple type does not generate messages at all. ===> at all during execution.

The principled type applies special mechanisms to adaptively decide whether to send the messages (equivalently, whether to \emph{prune} the messages), so the message quantity can be controlled to some extent. Our GACML is an instance of such method, and the most relevant studies include ATOC \cite{jiang2018learning}, MADDPG-M \cite{kilinc2018multi}, SchedNet \cite{kim2018learning} and IC3Net \cite{singh2018individualized}. Both ATOC and IC3Net learn a binary action to specify whether the agent wants to communicate with others, which is similar to our gating mechanism. However, ATOC and IC3Net are only suitable for homogeneous agents due to their DNN structures, while our GACML is a general model for both homogeneous and heterogeneous agents. More importantly, both ATOC and IC3Net cannot control how many messages will be pruned, while GACML has this ability because we introduce a special threshold for that purpose in our method. MADDPG-M adopts a two-level policy to learn whether the agent's private observation is sufficiently informative to be shared with others. However, MADDPG-M is only applicable to 2D tasks where the distance between the agents and their target agents is measurable and available. SchedNet generates a weight $w$ for each agent, and the top $M$ agents in terms of their weights $w$ will communicate with each other. However, it is not easy to apply a hard-coded parameter $M$ to real-world multi-agent systems. % Besides, ATOC only has a local range of communication, and it also suffers from the non-stationary problem. IC3Net prune the messages for private protection in competitive environments.

\textbf{Summary.} Most of the existing RL-based communication models are inflexible to be applied in the real-world multi-agent systems with limited-bandwidth restriction. A few possible DRL-based methods seem to be still preliminary, while our GACML is the first formal method that can control the message quantity to a desired threshold, as far as we know. % Although communication has long been an interest of AI community,

\section{The Proposed Method}
%We list the key variables used in this paper as follows. % in Table \ref{tab:variables}.
%\begin{table}[!htb]
%    \newcommand{\tabincell}[2]{\begin{tabular}{@{}#1@{}}#2\end{tabular}}
%    \centering
%    \begin{tabular}{|l|l|}
%        \hline
%        \bf $a_i$ & The local action of agent $i$. \\
%        \hline
%        \bf $\vec{a}_{-i}$ & The joint action of other agents except for agent $i$. \\
%        \hline
%        \bf $\vec{a}$ & The joint action of all agents, i.e., $\vec{a}=\langle a_{i}, \vec{a}_{-i} \rangle$. \\
%        \hline
%        \multicolumn{2}{|l|}{\tabincell{l}{The observation (history) $\vec{o}$, $o_i$, $\vec{o}_{-i}$, and the policy $\mu_{\theta_i}$ \\ and its parameters $\theta_i$ are denoted similarly.}}  \\
%        \hline
%        \hline
%        \bf $s'$ & The next state after $s$. \\
%        \hline
%        \multicolumn{2}{|l|}{$\vec{o}'$, $o_i'$, $\vec{o}_{-i}'$, $\vec{a}'$, $a_i'$, and $\vec{a}_{-i}'$ are denoted similarly.}  \\
%        \hline
%    \end{tabular}
%    %\caption{The key variables used in this paper.}
%    %\label{tab:variables}
%\end{table}

\subsection{ACML}
The key variables used in this section are as follows. $a_i$ is the local action of agent $i$. $\vec{a}_{-i}$ is the joint action of other agents except for agent $i$. $\vec{a}$ is the joint action of all agents, i.e., $\vec{a}=\langle a_{i}, \vec{a}_{-i} \rangle$. The observation history $\vec{o}$, $o_i$, $\vec{o}_{-i}$, and the policy $\mu_{\theta_i}$ are denoted similarly. $s'$ is the next state after $s$, and $\vec{o}'$, $o_i'$, $\vec{o}_{-i}'$, $\vec{a}'$, $a_i'$, $\vec{a}_{-i}'$ are denoted similarly.

\begin{figure}[!htb]
    \centering
    \includegraphics[height=6.5cm]{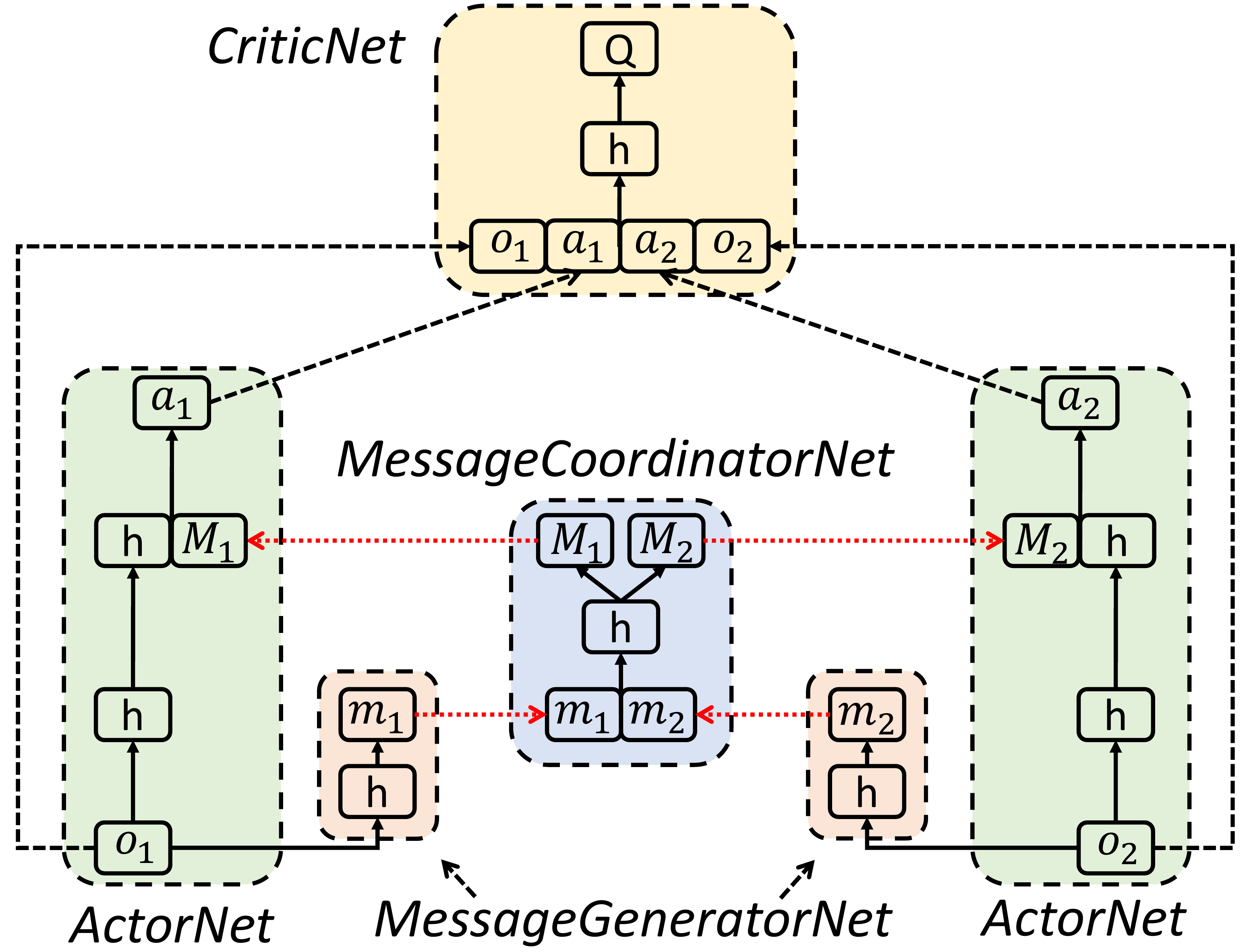}
    \caption{The proposed ACML. For clarity, we show this model using a two-agent example. All components are made up of DNN. $h$ is the hidden layer of the DNN; $m_i$ is the local message; $M_i$ is the global message. The red arrows imply the message sending process.}
    \label{fig:MODEL_ACML}
\end{figure}

\textbf{The Design.} ACML is motivated by combining the merits of the existing DRL-based communication methods mentioned in Section \ref{sec:RelatedWork}. As can be seen from Figure \ref{fig:MODEL_ACML}, ACML adopts the following designs.

(1) Each agent is made up of an ActorNet and a MessageGeneratorNet. This design represents the policy and the message with two separated network branches. Another choice is to take the hidden layer of policy network as the message. We notice that most previous methods adopt the former, which usually outperforms the latter.

(2) All agents share the same CriticNet and MessageCoordinatorNet, which are placed in a \textbf{\emph{coordinator}} (i.e., a specially designed agent). The shared MessageCoordinatorNet is similar to many previous methods such as the CommNet, AMP and ATOC, while the shared CriticNet is similar to the well-known MADDPG and COMA.

Although each separate design is common, aggregating them together  properly is novel. For example, MADDPG and COMA do not adopt the MessageGeneratorNet and MessageCoordinatorNet, making them suffer from the partially observable problem during execution; while AMP and ATOC do not adopt the shared CriticNet, making them suffer from the non-stationary problem \cite{hernandez2017survey} during training. ACML is fully observable and training stationary. %, theoretically.

ACML works as follows during execution \footnote{Please note that the CriticNet is only used during training, while other components are needed during execution.}. % as shown in Figure \ref{fig:MODEL_ACML}.

(1) $m_i = \text{MessageGeneratorNet}(o_i)$, i.e., agent $i$ generates the local message $m_i$ based on its own observation $o_i$.

(2) All agents send their $m_i$ to the coordinator.

(3) $M_1,.., M_N = \text{MessageCoordinatorNet}(m_1,.., m_N)$, i.e., the coordinator extracts the global message $M_i$ for each agent $i$ based on all local messages $m_i$.

(4) The coordinator sends $M_i$ back to agent $i$.

(5) $a_i = \text{ActorNet}(o_i, M_i)$, i.e., agent $i$  generates action $a_i$ based on its local observation $o_i$ and the global message $M_i$, which  encodes all $\langle o_1,.., o_N \rangle$ for full observability.

\textbf{The Training.} As described above, the agents generate $a_i$ based on $o_i$ and $M_i$ to interact with the environment, and the environment will feed a shared reward $r$ back to the agents. Then, the experience tuples $\langle o_i, \vec{o}_{-i}, a_i, \vec{a}_{-i}, r, o'_i,\vec{o}'_{-i} \rangle$ are used to train ACML. Specifically, as the agents exchange messages with each other, the actor and the shared critic can be represented as $\mu_{\theta_i}(o_i,M_i)$ and $Q(\vec{o},\vec{a};w)$, respectively. We can extend Equation (\ref{equ:DPG1} -- \ref{equ:DPG3}) to multi-agent formulations shown in Equation (\ref{equ:mDPG1} -- \ref{equ:mDPG3}), where the parameters $w$, $w^{-}$, $\theta_i$ and $\theta_i^{-}$ have similar meaning to these of single-agent setting.
\begin{eqnarray}
    \delta =& \hspace{-0.8em} r+\gamma Q(\vec{o}', \vec{a}';w^{-})|_{a'_i=\mu_{\theta_i^{-}}(o'_i)} - Q(\vec{o}, \vec{a};w) \label{equ:mDPG1} \\
    L(w) =& \hspace{-6.3em} \mathbb{E}_{(o_i,\vec{o}_{-i},a_i,\vec{a}_{-i},r_i,o'_i,\vec{o}'_{-i}) \sim D}[\delta^{2}] \label{equ:mDPG2} \\
    \nabla_{\theta_i}J(\theta_i) =& \hspace{-6.8em} \mathbb{E}_{(o_i,\vec{o}_{-i}) \sim D}[\hspace{0.3em} \nabla_{\theta_i}\mu_{\theta_i}(o_i, M_i)  \nonumber \\
    & \hspace{3.8em} * \hspace{0.2em}  \nabla_{a_i}Q(\vec{o}, \vec{a};w)|_{a_i=\mu_{\theta_i}(o_i)} \hspace{0.2em} ] \label{equ:mDPG3}
\end{eqnarray}

Since all components in ACML are implemented by DNN, ACML is end-to-end differentiable, and the communication message and the control policy can be optimized jointly using back propagation (BP) based on the above equations.

\subsection{GACML} \label{sec:GACML}
\textbf{Motivation.} As can be seen from the execution process, ACML takes as input the communication messages from \emph{all} agents to generate \emph{a single} control action. Thus, the agents have to keep sending messages incessantly in every control cycle, regardless whether the messages are beneficial enough to the performance of the agent team. This is also a common problem of most deep communication models as mentioned in Section \ref{sec:RelatedWork}. As a result of too many messages, these methods are inflexible to be applied to the real-world multi-agent systems with limited-bandwidth restriction.

GACML is motivated by handling this problem through pruning unprofitable messages in a principled manner. % and dynamic manner.

\begin{figure}[!htb]
    \centering
    \includegraphics[height=4.2cm]{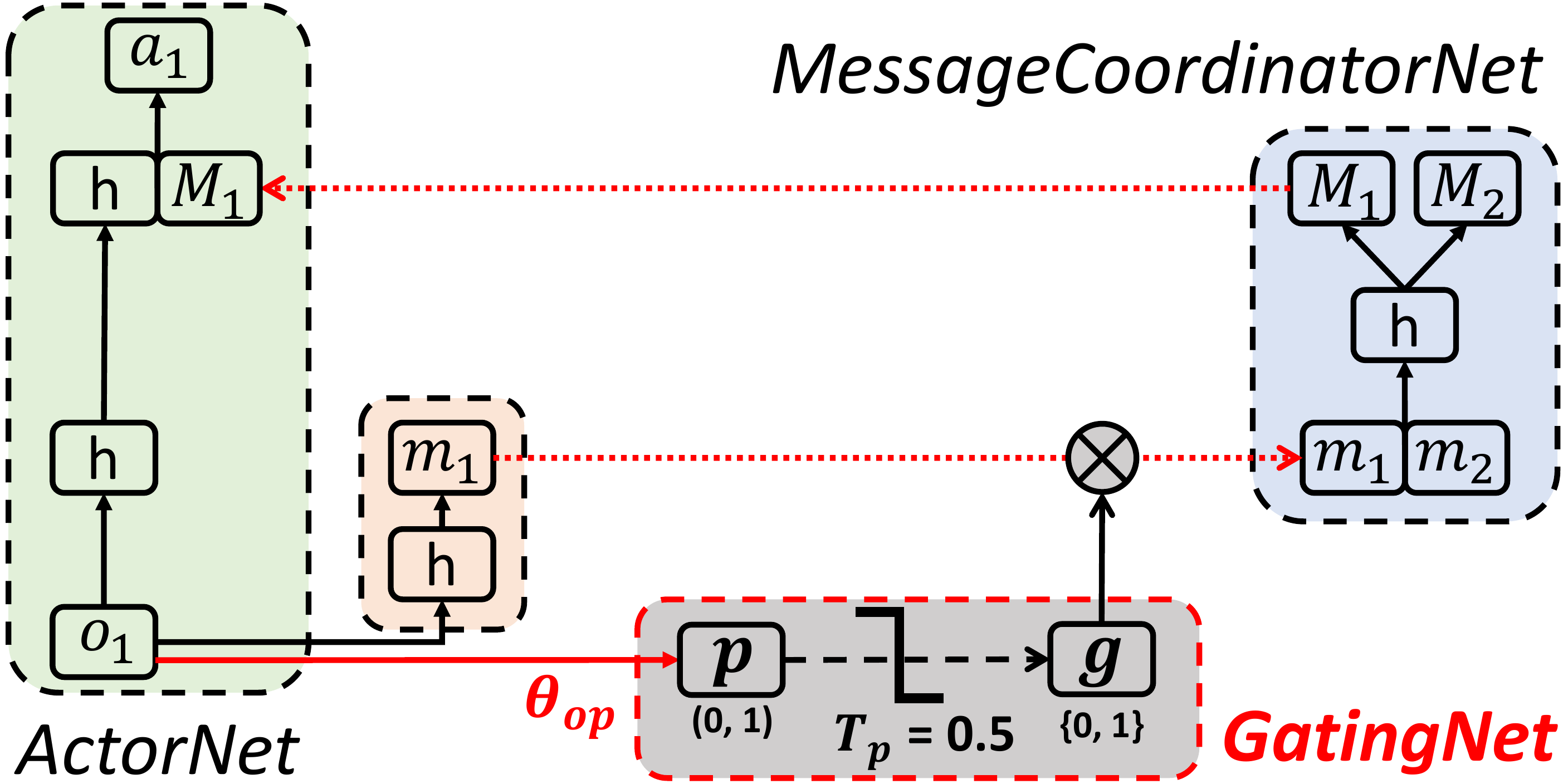}
    \caption{The actor part of the proposed GACML. For clarity, we only show one agent's structure, and we do not show the critic part because it is the same as that of ACML.}
    \label{fig:MODEL_ACML_Gated}
\end{figure}

% \textbf{The Design.} We propose a gating mechanism to adaptively prune unprofitable messages among ActorNets \footnote{The critic will be naturally removed during execution.}, such that the agents can generate coordinated actions even without getting access to these messages. % This more flexible model is named as GACML.
\textbf{The Design.} We propose a gating mechanism to adaptively prune unprofitable messages among ActorNets, such that the agents can maintain the performance while pruning as many messages as possible to resolve the bandwidth constraints.

As shown in Figure \ref{fig:MODEL_ACML_Gated}, except for the original components, each agent is equipped with an additional GatingNet. Specifically, GACML works as follows.

(1) The agent generates a local message $m_i$ as well as a probability score $p \in (0, 1)$ based on its own observation $o_i$.

(2) A gate value $g \in \{0, 1\}$ is generated by a threshold function $T_{p}=0.5$. That is to say, if $p \le T_p$, we set $g=0$, otherwise, we set $g=1$.

(3) The agent sends $m_i \odot g$ to the coordinator, and the following process is the same as that of ACML.

In the above process, if $g=0$, $m_i \odot g$ will be a zero vector, and the agent has no need to send $m_i$ to the coordinator. Accordingly, the coordinator does not send the global message $M_i$ back to the agent, because we replace $M_i$ with a zero vector to ``tell'' our GACML model that the message $M_i$ is pruned indeed. We call this design \emph{zero padding}.

Please note that zero padding is a crucial design for pruning a lot of messages. Suppose that we have pruned many messages, the time span of two valid messages would be relatively large. If we simply adopt the message caching method \cite{dowling2008decentralized} to replace $M_i^{t}$ with $M_i^{\tilde{t}}$ generated a long time ago (i.e., $t - \tilde{t}$ is relatively large), $M_i^{\tilde{t}}$ can hardly approximate $M_i^{t}$ anymore. In contrast, the zero padding will always ``tell'' GACML a correct signal: the messages have been really pruned. Using another GatingNet to decide whether to prune $M_i$ will be evaluated in the future.

% An alternative way is to use another GatingNet to decide whether to prune $M_i$. This design is left for future evaluation. % We call this message pruning\footnote{Alternatively, one can use another GatingNet to decide whether to prune $M_i$. Here, we simply adopt the message caching method.}.  , and it will confuse the network about whether $M_i$ is pruned indeed

% Note that we cannot use a soft gate, e.g., sending $m \cdot p$ to the coordinator. On one hand, this will change the scale of $m$. On the other hand, $m \cdot p$ should always be sent no matter how small $p$ is, which cannot reduce the number of messages.

\textbf{The Training Method with a Novel Auxiliary Task.}
In order to make the above design work, a suitable probability $p$ must be trained for each observation $o$ \footnote{We remove the subscript $i$ to represent that the following description is suitable for all agents.}, otherwise GACML may degenerate to ACML (in the extreme case where $p$ is always larger than $T_p$). However, as the threshold function $T_p$ is non-differentiable, it makes the end-to-end BP method inapplicable. We also tried the approximate gradient \cite{Hubara2016Binarized}, the sparse regularization \cite{Makhzani2015Winner} and several other methods without success. % but without success. A possible reason is that the gradient is originally coming from the TD-error computed at the critics, which is so far away from $p$ that correct training signal is dissipated, so we decide to embrace the auxiliary task technique ...

Finally, we decided to embrace the auxiliary task technique \cite{Jaderberg2016Reinforcement} to provide training signal for $p$ directly. Recall that we want to prune the messages on the premise of maintaining the performance. For RL, the performance could be measured by the Q-value, so we design the following auxiliary task.

Let $p$ indicate the probability of that $\Delta Q(o) = Q(o,a^{C}) - Q(o,a^{I})$ is larger than $T$, where $a^{C}$ is the action generated based on communication, $a^{I}$ is the action generated independently (i.e., without communication), and $T$ is a threshold controlling how many messages should be pruned. In this setting, the label of this auxiliary task can be formulated as
\begin{eqnarray}
    Y(o) = \mathbb{I}(Q(o,a^{C}) - Q(o,a^{I}) > T) \label{equ:AuxiliaryLabel}
\end{eqnarray}
where $\mathbb{I}$ is the indicator function. Then we can train $p$ by minimizing the following loss function:
\begin{eqnarray}
    L_{\theta_{op}}(o) = - \mathbb{E}_o[Y(o) \log p(o;\theta_{op}) \nonumber \\
    & \hspace{-8.2em} + (1-Y(o)) \log (1-p(o;\theta_{op}))] \label{equ:gating}
\end{eqnarray}
where $\theta_{op}$ are the parameters between the observation $o$ and the probability $p$ as shown in Figure \ref{fig:MODEL_ACML_Gated}.

The insight of the above loss function is that if $\Delta Q(o) = Q(o,a^{C}) - Q(o,a^{I})$ is really larger than $T$ (i.e., $a^{C}$ can obtain at least $T$ Q-values more than $a^{I}$, and the corresponding label is $Y(o)=1$), the network should try to generate a probability $p(o;\theta_{op})$ that is larger than $T_p=0.5$ to encourage communication. In other word, GACML only prunes messages that contribute less Q-values than the threshold $T$. Therefore, after the gating mechanism is trained well, it can prune unprofitable messages adaptively to control the message quantity around a desired threshold specified by $T$. Consequently, GACML needs much fewer messages to achieve a desirable performance. It is our key contribution that
makes GACML a novel and principled method. % , and it is applicable to real-world tasks with limited-bandwidth restriction.

\textbf{The Key Implementation.}
The above training method relies on correct labels of the auxiliary task. It means that we should provide suitable $Q(o,a^{C})$, $Q(o,a^{I})$ and $T$ as indicated by Equation (\ref{equ:AuxiliaryLabel}).

For the Q-values, we firstly set $g=1$ (i.e., without message pruning) to train other components except for the GatingNet based on Equation (\ref{equ:mDPG1} -- \ref{equ:mDPG3}). After the model is trained well, we can get approximately correct ActorNet and CriticNet. Afterwards, for a specific observation $o$, the ActorNet can generate $a^{C}$ and $a^{I}$ when we set $g=1$ and $g=0$, respectively; then the CriticNet can estimate an approximately correct Q-value $Q(o,a^{C})$ and $Q(o,a^{I})$.

For the threshold $T$, we propose two methods to set a fixed $T$ and a dynamic $T$, respectively. To calculate a fixed $T$, we firstly sort the $\Delta Q(o)$ of the latest $K$ observations $o$ encountered during training, resulting in a list of $\Delta Q(o)$, which is called $L_{\Delta Q(o)}$. Then, we set $T$ by splitting $L_{\Delta Q(o)}$ in terms of the index. For example, if we want to prune $T_m\%$ messages, we set $T=L_{\Delta Q(o)}[K \times T_m\%]$. We do not split $L_{\Delta Q(o)}$ in terms of the value, since $\Delta Q(o)$ usually has a non-uniform distribution. The advantage of a fixed $T$ is that the actual number of the pruned messages is ensured to be close to the desired $T_m\%$. Besides, this method is friendly to a large $K$.

For the dynamic $T$, we adopt the exponential moving average technique \footnote{\url{https://en.wikipedia.org/wiki/Moving_average\#Exponential_moving_average}} to set $T$:
\begin{eqnarray}
    T_{t} = (1-\beta) T_{t-1} + \beta (Q(o_t,a^{C}_t) - Q(o_t,a^{I}_t)) \label{equ:T}
\end{eqnarray}
where $\beta$ is a coefficient for discounting older $T$. We test some $\beta$ in $[0.6,0.9]$, and they all work well. The advantage of a dynamic $T$ is that $Y(o)$ becomes an adaptive training label even for the same observation $o$. This is very important for the dynamically changing environments, because $T$ and $Y(o)$ can quickly adapt to these environments.

% In practice, we train the GACML by alternately executing two steps: first, training other components based on Equation (\ref{equ:mSPG1} -- \ref{equ:mDPG3}) to get more correct ActorNet and CriticNet, and consequently more correct $a^{C}$ and $a^{I}$, $Q(o,a^{C})$ and $Q(o,a^{I})$, threshold $T$ and label $Y(o)$; second, training the probability $p$ based on Equation (\ref{equ:gating}). Alternately training can adapt to the dynamic environments more easily.

\section{Experiments}
\subsection{The Experimental Settings}
Due to limited space, we only show the evaluation on real-world packet routing systems. Four benchmark tasks such as traffic control and predator prey are shown in the appendix. % \footnote{Anonymous appendix \url{https://anonymousfiles.io/OtyM1Qlh/}}.
% \footnote{The anonymous appendix: \url{https://dwz.cn/wVmuaCNY}.}.

% We adopt Independent Actor-Critic (IND-AC) \cite{peng2017multiagent}, MADDPG \cite{lowe2017multi}, and AMP \cite{peng2018learning} for comparison. They can be seen as the ablation models of our ACML. In IND-AC, the agent learns its own actor-critic network independently without communication. We can know the effect of communication by comparing with IND-AC. MADDPG adopts centralized critics to share information among multiple agents, while the actors are independent. Therefore, MADDPG suffers from partially observable problem during execution. In AMP, only the actors exchange messages, while the critics are independent. Thus, AMP suffers from non-stationary problem during training. In contrast, both actors and critics in ACML can exchange messages, so that these problems are expected to be relieved. Methods like ATOC are not compared, since they are inapplicable to the routing tasks due to the reasons analysed in Section \ref{sec:RelatedWork}. %, without essential modification.

\textbf{Baseline.} We adopt Independent Actor-Critic (IND-AC) \cite{peng2017multiagent}, MADDPG \cite{lowe2017multi} and AMP \cite{peng2018learning} for comparison. These methods can be regarded as the ablation models of ACML. In IND-AC, the agent learns its own actor-critic network independently without communication. We can know the effect of communication by comparing with IND-AC. MADDPG adopts centralized critics to share information among multiple agents, while the actors are independent. In AMP, only the actors exchange messages, while the critics are independent. In contrast, both actors and critics in ACML can exchange messages. Methods like ATOC are not compared, since they are unsuitable for the routing tasks due to the reasons analyzed in Section \ref{sec:RelatedWork}. %, without essential modification.

 % Please note that only IND-AC, MADDPG and GACML are applicable to systems with strict limited-bandwidth restriction, because AMP keeps sending messages during execution, resulting too many messages.

% \textbf{Parameter.} The neural network structure is the same as that of Figure \ref{fig:MODEL_ACML} and \ref{fig:MODEL_ACML_Gated}. The probability $p$ is a hidden layer with 1 neuron, which generates a scaler value with a Sigmoid activation function. The first hidden layer has 64 neurons, while other hidden layers have 32 neurons, and the activation function of these layers is Relu. Other hyperparameters are as follows: the learning rates of actor, critic and target networks are 0.001, 0.01 and 0.001, respectively; replay buffer size is $10^6$; batch size is 128; discount factor is 0.95.
\textbf{Parameter.} The probability $p$ in GACML is a hidden layer with 1 neuron, and the activation function is Sigmoid. The first hidden layer of our model has 64 neurons, while other hidden layers have 32 neurons, and the activation function is Relu. Learning rates of actor, critic and target networks are 0.001, 0.01 and 0.001, respectively. Replay buffer size, batch size and discount factor are $10^6$, 128 and 0.95, respectively.

\subsection{The Packet Routing System}
\begin{figure}[!htb]
    \centering
    \subfigure[The small topology.]{
        \label{fig:twoIE}
        \includegraphics[height=2.2cm,width=3.8cm]{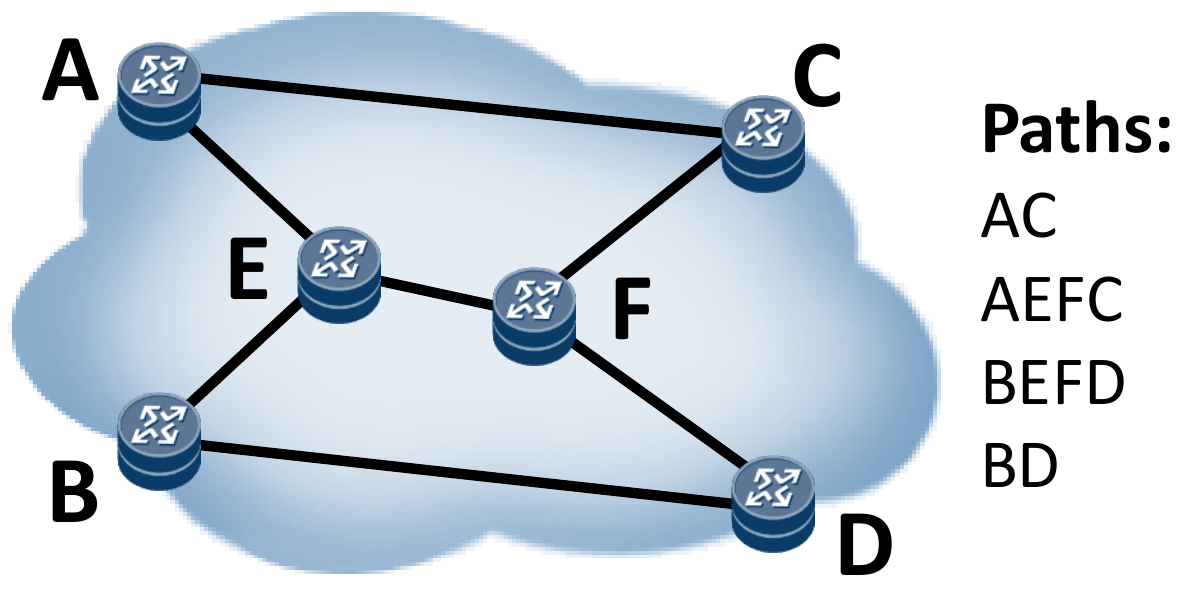}
    }
    \hspace{0.0cm}
    \subfigure[The moderate topology.]{
        \label{fig:fiveIE}
        \includegraphics[height=2.2cm,width=4.2cm]{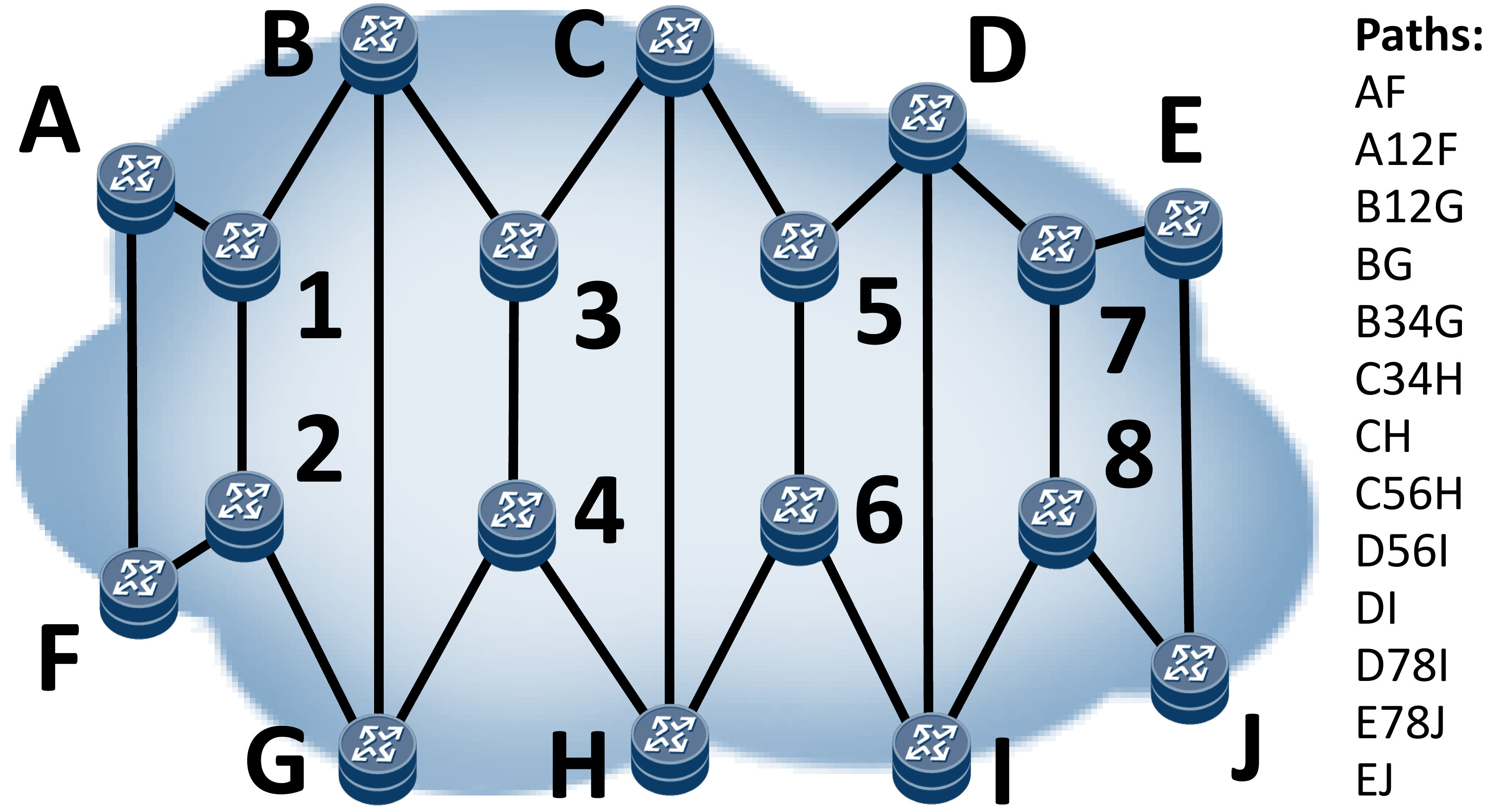}
    }
    \hspace{0.05cm}
    \subfigure[The large topology.]{
        \label{fig:WAN}
        \includegraphics[height=2.8cm]{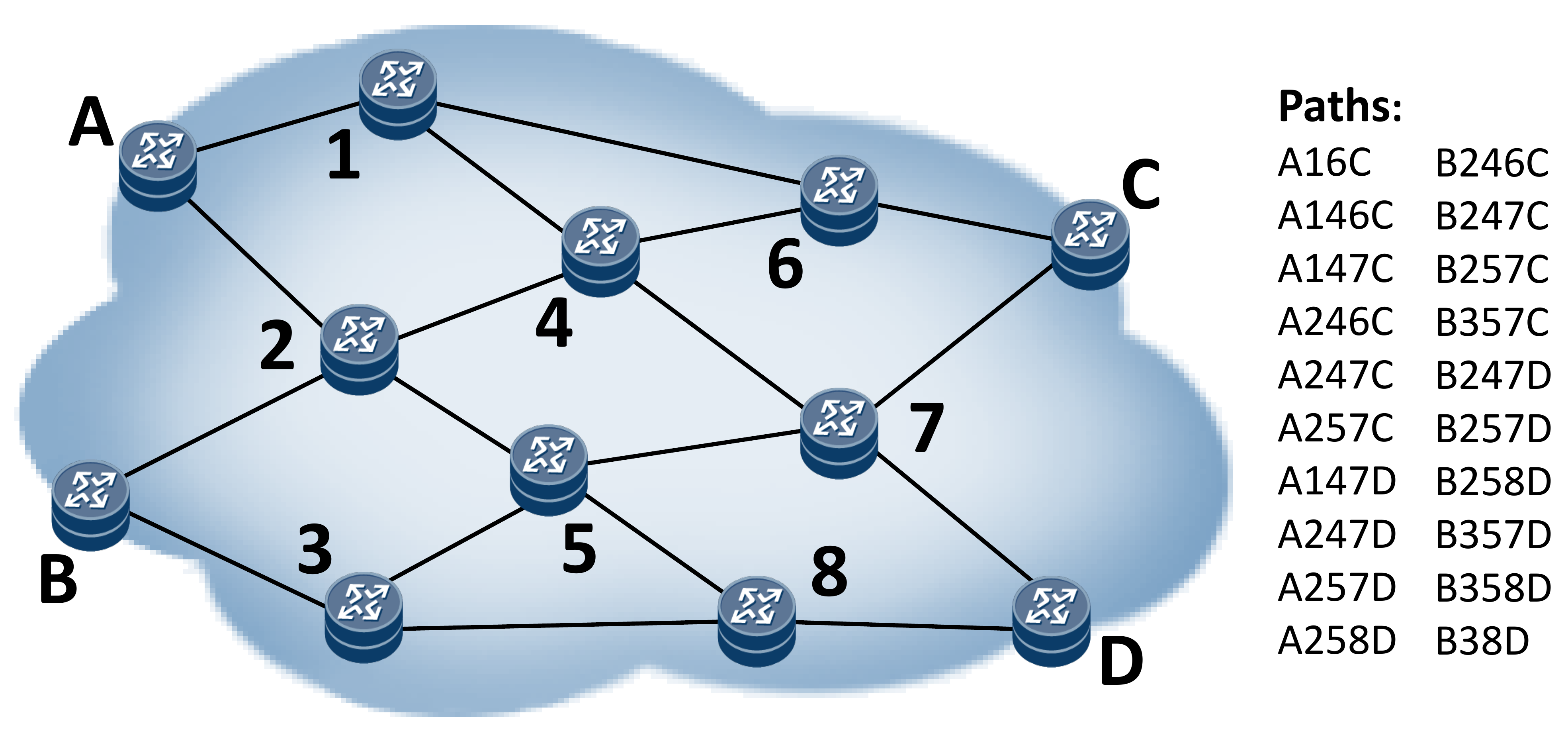}
    }
    \caption{The packet routing environments.}
    \label{fig:PacketRoutingEnvironments}
\end{figure}

\textbf{Environment Description.} In the information era, packet routing is a very fundamental and critical task on the Internet. We evaluate our methods on three routing tasks. As shown in Figure \ref{fig:PacketRoutingEnvironments}, the small topology and the moderate topology are the most classical topologies in the Internet Traffic Engineering community \cite{Kandula2005Walking}; the large topology is based on the real needs of our industrial collaborator, and it is more complex than the real-world Abilene Network \footnote{A backbone net \url{https://en.wikipedia.org/wiki/Abilene_Network}} in terms of the numbers of routers, links and paths.

In each topology, there are several edge routers. Each edge router has an aggregated flow that should be transmitted to other edge routers through available paths. For example, in Figure \ref{fig:twoIE}, $B$ is set to transmit flow to $D$, and the available paths are $BEFD$ and $BD$. Each path is made up of several links, and each link has a \emph{link utilization}, which equals to the ratio of the current flow on this link to the maximum flow transmission capacity of this link.

The necessity of cooperation among routers is as follows: one link can be used to transmit the flow from more than one router, so the routers must not split too much or too little flow on the same link at the same time; otherwise this link will be either overloaded or underloaded.

Note that there is a strict limited-bandwidth restriction in these tasks, because the capacity of the network is limited. %, and the routers should try to transmit more flow packets rather than the communication messages.

\textbf{Problem Definition.} The routers are controlled by our algorithm, and they try to learn a good flow splitting policy to minimize the \emph{Maximum Link Utilization in the whole network (MLU)}. The intuition behind this goal is that high link utilization is bad for dealing with bursty flow. The \textbf{\emph{observation}} includes the latest ten steps' flow demands,
the latest ten steps' link utilizations and their average, and the latest
action taken by the router. The \textbf{\emph{action}} is a flow splitting ratio on each available path (e.g., in Figure \ref{fig:fiveIE}, B will generate an action like $[a\%, b\%, 1-a\%-b\%]$ for paths $[B12G, BG, B34G]$). The \textbf{\emph{reward}} is set as $1-MLU$ to minimize \emph{MLU}. Besides \emph{MLU}, we also care about the \emph{convergence ratio} of all experiments. % Exploration bonus based on local link utilization can be added accordingly. \footnote{The detailed advantages of minimizing MLU is discussed in \cite{Kandula2005Walking}.}

\subsection{The Experimental Results}
\textbf{Results without Message Pruning.} In this experiment, we use synthetic flow to evaluate different methods on the small and moderate topologies. The flow has a shape of $A\sin(wx+\varphi)+b$ with different settings of $A$, $w$, $\varphi$, $b$.

\begin{figure}[!htb]
    \centering
    \subfigure[The MLU.]{
        \label{fig:RESULT_PacketRouting_Preliminary_MLU}
        \includegraphics[height=2.9cm,width=3.3cm]{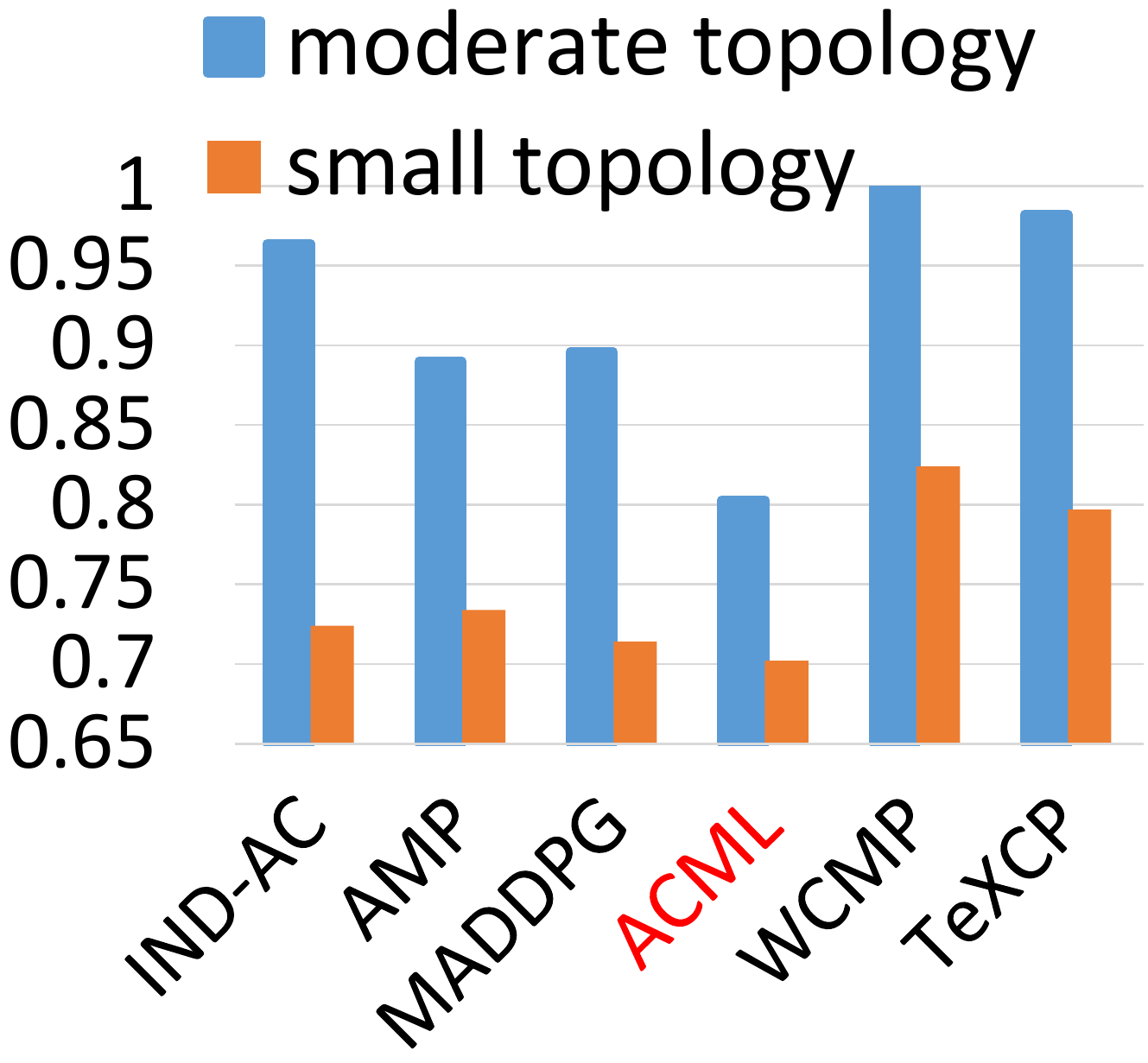}
    }
    \hspace{0.5cm}
    \subfigure[The convergence ratio.]{
        \label{fig:RESULT_PacketRouting_Preliminary_ConvergenceRate}
        \includegraphics[height=2.9cm,width=3.3cm]{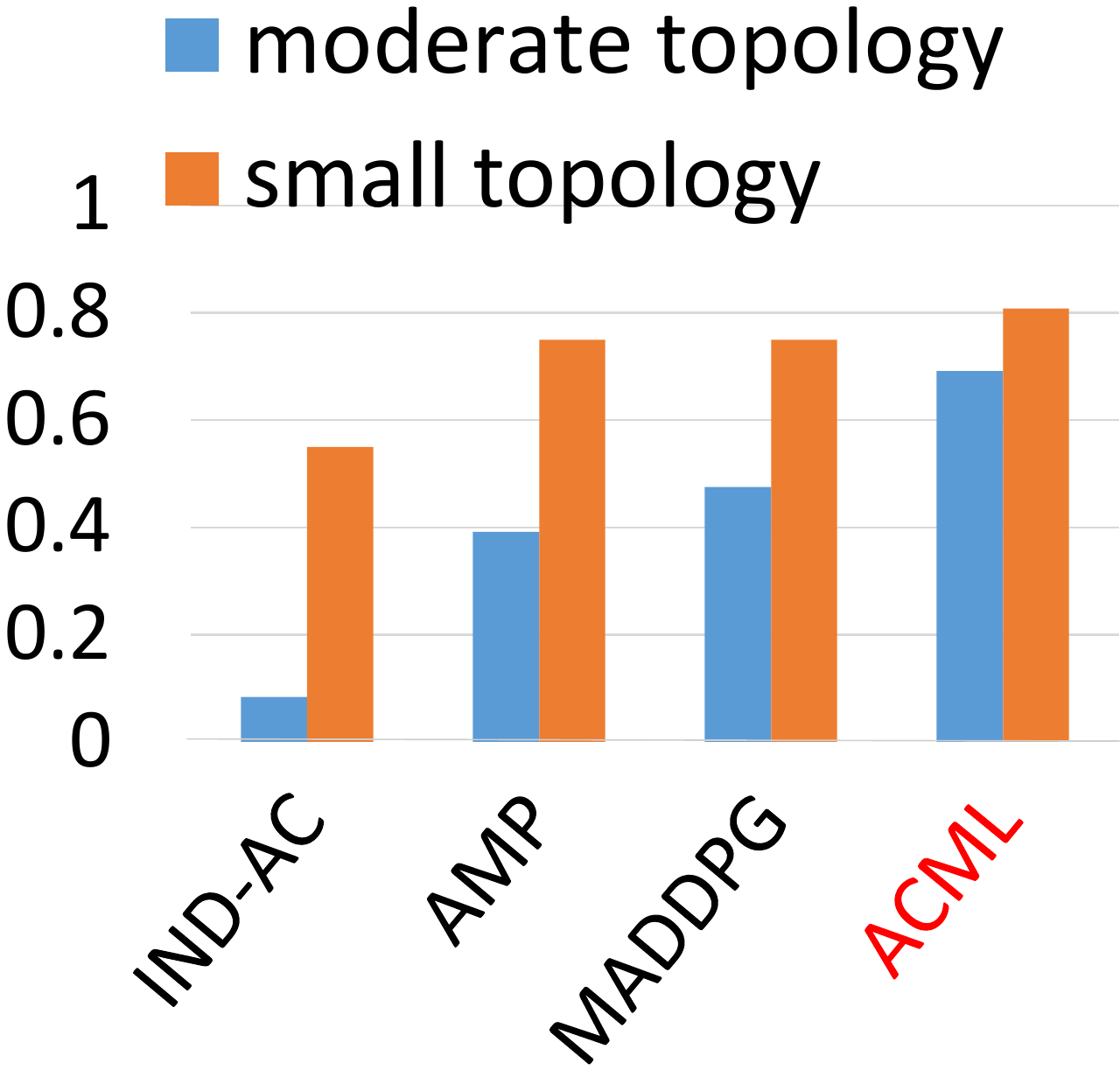}
    }
    \caption{The average results of 30 independent experiments. A smaller MLU is better. A larger convergence ratio is better.}
    % MLU is the max while convergence rate is average of all 30 experiments
    \label{fig:RESULT_PacketRouting_Preliminary}
\end{figure}

Figure \ref{fig:RESULT_PacketRouting_Preliminary_MLU} shows the MLU of 30 independent experiments. In the small topology, all DRL-based methods (i.e., excluding the rule-based WCMP and TeXCP) have a similar performance. The reason is that this topology is rather simple, and all DRL-based methods can find the near-optimal control policies easily after they have been trained, regardless whether the methods adopt communication. Thus, the more advanced methods, such as MADDPG and ACML, do not have enough space to further improve IND-AC.

In the moderate topology, the performances of IND-AC, AMP and MADDPG drop severely, while ACML keeps the performance and achieves much smaller MLU. This is because other models suffer from either the partially observable problem or the non-stationary problem, making them ineffective in complex environment. In contrast, ACML can address these problems as analyzed before, resulting in stronger ability to deal with complex tasks.

Furthermore, compared with the rule-based TeXCP \footnote{We slightly modify it for the testing environment.} \cite{Kandula2005Walking} and WCMP \footnote{\url{https://en.wikipedia.org/wiki/Equal-cost_multi-path_routing}}, the DRL-based models achieve much smaller MLU in both topologies. The reason lies in that the DRL-based methods can take the actions' future effect into consideration, which is in favor of accomplishing the routing task, whereas the rule-based methods can only consider the current effect of the actions. Even worse, WCMP gets a MLU larger than 1.0 in the moderate topology, and the simulation system crashed.

Figure \ref{fig:RESULT_PacketRouting_Preliminary_ConvergenceRate} shows the convergence ratio. As can be seen, ACML achieves the largest convergence ratio in both topologies. Furthermore, when the evaluation turns to the moderate topology, the decrease of convergence ratio of ACML is much smaller than that of other methods. It means that our ACML is more stable than other DRL-based methods during training.

Overall, the above results demonstrate ACML with general applicability, scalability and stability.

\textbf{Message Pruning Evaluation.} In this experiment, we test GACML using real flow trajectory and the large topology.

The results for adopting a fixed threshold $T=L_{\Delta Q(o)}[K \times T_m\%]$ are shown in Table \ref{tab:RESULT_Message Pruning_Test}. We draw the following conclusions. (1) For a predefined threshold $T_m\%$, the actual number of the pruned messages is close to $T_m\%$. It means that GACML can control the message to our desired quantity. (2) GACML can prune a large number of messages (e.g., more than 82\%) with little damage to the performance (e.g., less than 7\%). It implies that the pruned messages are usually not beneficial enough to the performance. (3) When we pruned all messages (i.e., the last row of Table \ref{tab:RESULT_Message Pruning_Test}), the reward has a large decrease. Compared with the second to the last row of Table \ref{tab:RESULT_Message Pruning_Test}, it indicates that the remaining 1.5\% messages are very important for keeping the performance, and GACML has learnt to share these important messages with all agents. %  retain and

\begin{figure}[!htb]
    \centering
    \includegraphics[height=3.2cm,width=7.5cm]{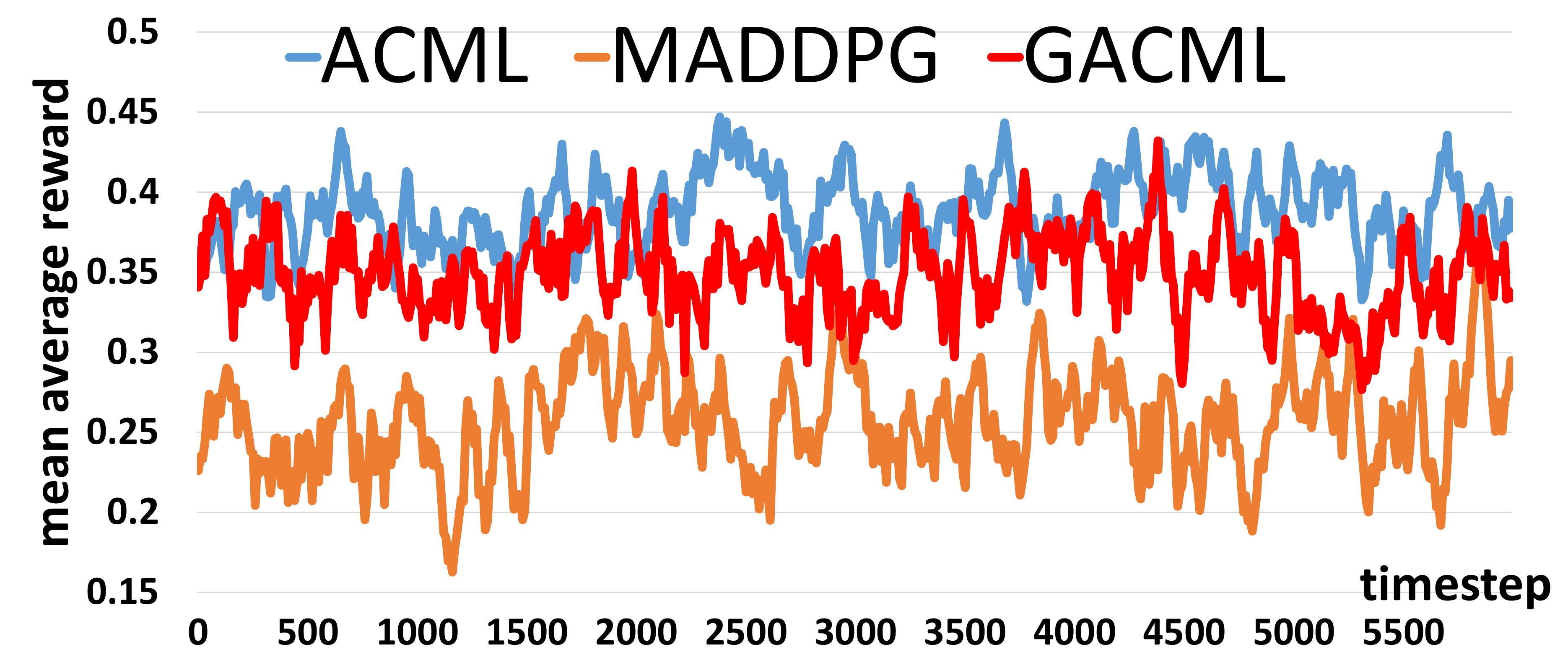}
    \caption{The testing rewards. For GACML, we set $\beta=0.8$. }
    \label{fig:WANgating_reward_realFlow}
\end{figure}

The results for adopting a dynamic threshold $T$ are shown in Figure \ref{fig:WANgating_reward_realFlow}. As can be seen, GACML performs much better than the state-of-the-art MADDPG. In addition, GACML can keep the performance close to ACML while pruning quite a lot of messages (in this case, 74.3\%).

To conclude, the above results demonstrate that the proposed gating mechanism can prune unnecessary messages on the premise of maintaining the performance indeed, and that our GACML is very suitable for the real-world routing systems with limited-bandwidth restriction.

\begin{table}[!htb]
    \centering
    \begin{tabular}{lrr}
        \toprule
        \bf $\bm{T_m}$\% & \bf \# of Pruned Message & \bf Reward Decrease \\
        \midrule
        \bf 80\% & 82.13\% & 6.84\% \\
        \bf 90\% & 87.68\% & 8.05\% \\
        \bf 95\% & 93.39\% & 11.30\% \\
        \bf 98\% & 99.14\% & 15.98\% \\
        \bf 99\% & 98.53\% & 14.86\% \\
        \bf 100\% & 100.00\% & 61.52\% \\
        \bottomrule
    \end{tabular}
    \caption{The pruned message and the reward decrease in terms of ACML for a fixed (i.e., a predefined) pruning threshold $T_m\%$.}
    \label{tab:RESULT_Message Pruning_Test}
\end{table}

\textbf{Message Pruning Analysis.} In this experiment, we analyze GACML using synthetic flow and the small topology.

\begin{figure}[!htb]
    \centering
    \subfigure[The average rewards.]{
        \label{fig:twoIEgating_reward}
        \includegraphics[height=1.8cm,width=3.9cm]{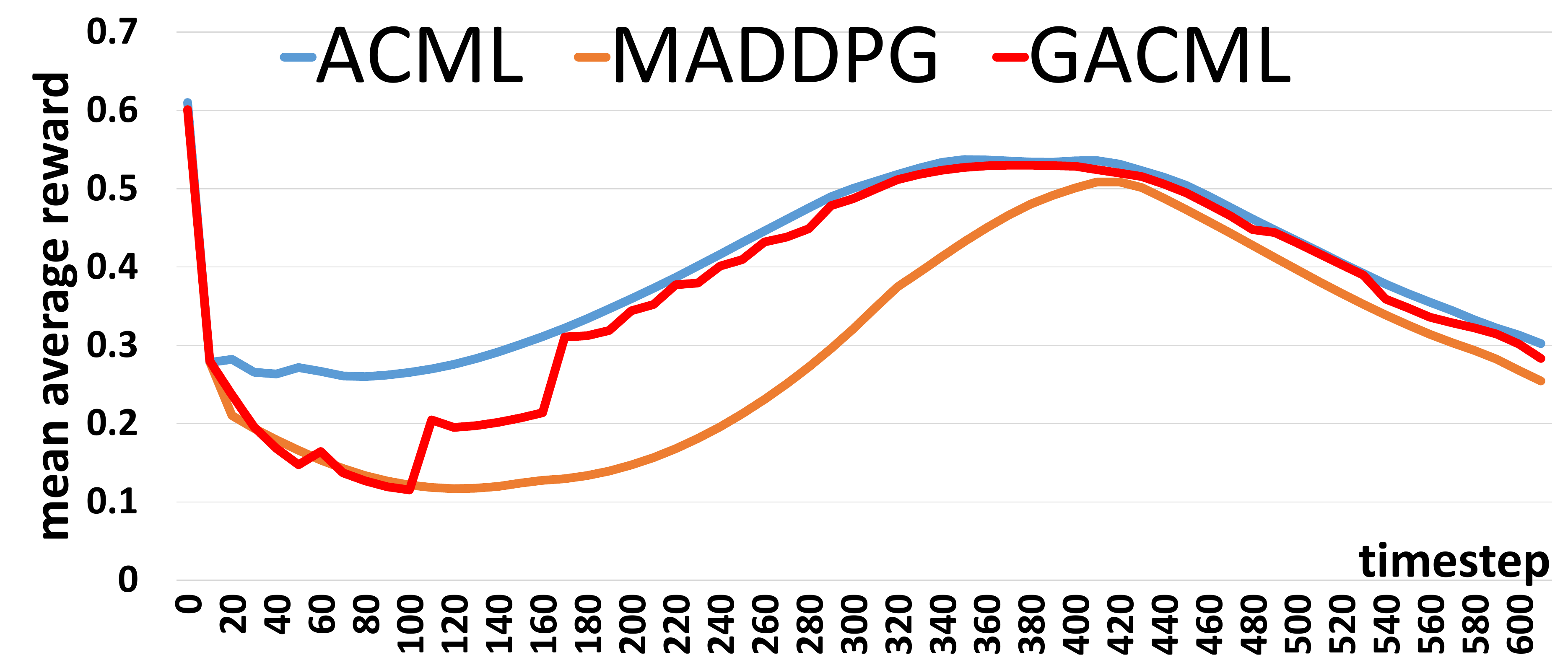}
    }
    \hspace{0.0cm}
    \subfigure[The gate values of 2 tests.]{
        \label{fig:twoIEgating_value}
        \includegraphics[height=1.8cm,width=3.9cm]{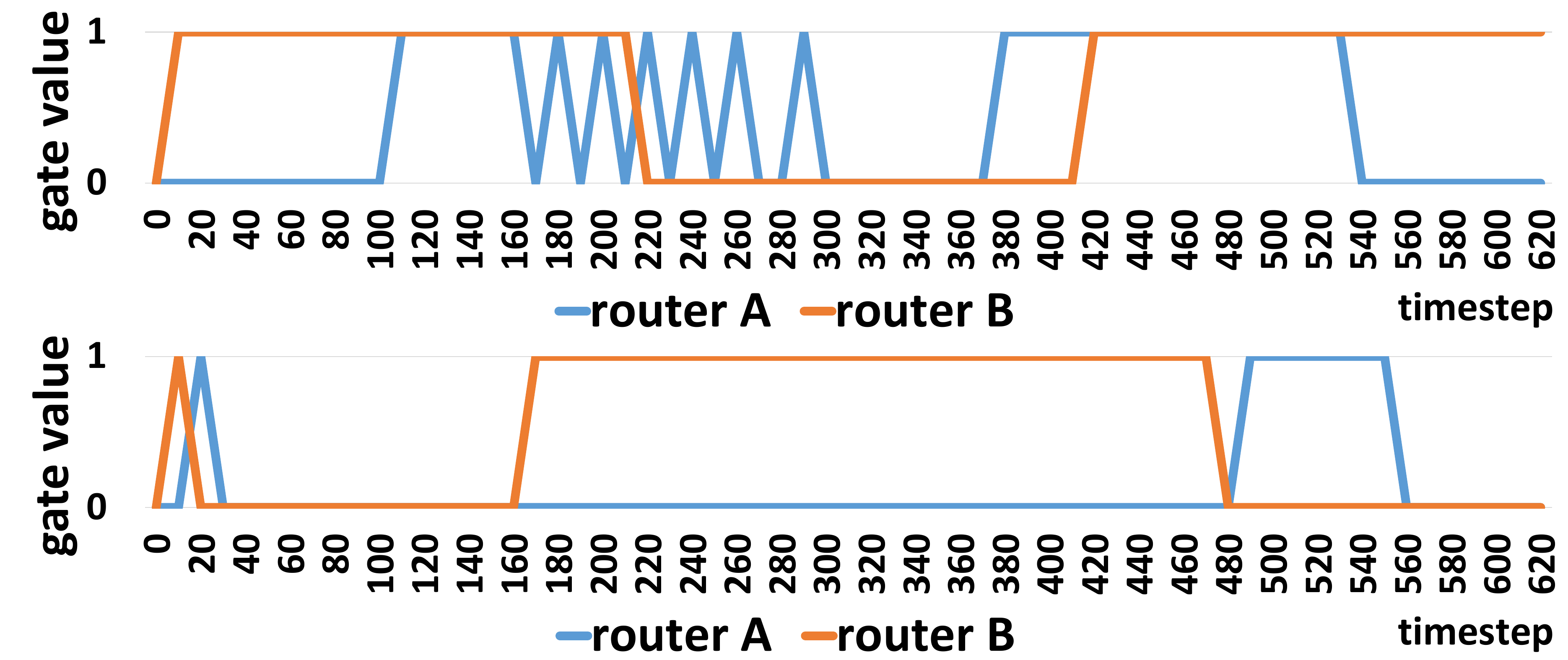}
    }
    \caption{The testing results of 30 message pruning experiments based on the small topology and synthetic flow. GACML gets 8.2\% fewer rewards than ACML, while the message decrease is 56.3\%. It also outperforms MADDPG. Please zoom in for better view.}
    \label{fig:twoIEgating}
\end{figure}

We adopt this setting because the results are easy to understand. For example, as shown in Figure \ref{fig:twoIEgating_reward}, when the total flow in the network is decreasing during timestep 100 to 300, GACML adapts to this situation more quickly than MADDPG, and it obtains similar rewards as ACML after dozens of steps. Besides, since we use $A\sin(wx+\varphi)+b$ to generate synthetic flow, the reward curves shown in Figure \ref{fig:twoIEgating_reward} and the gate value curves shown in Figure \ref{fig:twoIEgating_value} are similar to the shape of $\sin(x)$. We also notice that the bursty flow usually generates a gate value $g=1$. The results imply that GACML has learnt a subtle communication strategy.

\section{Conclusion}
This paper presents a method to jointly learn the communication strategy and the control policy for multiple cooperative agents. In contrast to most previous methods, we focus on addressing the limited-bandwidth restriction that exists in many real-world applications. Specifically, we firstly aggregate the merits of the existing methods to form a new method that outperforms several state-of-the-art DRL-based and rule-based methods. Then, we propose a gating mechanism with several crucial designs that can prune quite a lot of unprofitable messages with little damage to the performance. Consequently, our method is applicable to the real-world packet routing systems with limited bandwidth. As far as we know, it is the first formal method to achieve this in a novel and principled way, and it is the key contribution of this work. % It is our key contribution that makes GACML a novel and principled method.  in both the real-world packet routing tasks and the benchmark tasks

%% The file named.bst is a bibliography style file for BibTeX 0.99c
\bibliographystyle{named}
\bibliography{ijcai19}

\end{document}